\documentstyle[prl,aps,epsfig]{revtex}
\begin{document}
\draft
\twocolumn[\hsize\textwidth\columnwidth\hsize\csname@twocolumnfalse%
\endcsname

%\preprint{\parbox[t]{45mm}{\small  ADP-01 04 T439}}

\hfill{\small  ADP-01 04 T439}

\hfill{\small  FAU-TP3-01/4}

\title{On Domain-like Structures in the QCD Vacuum } 

\author{Alex C. Kalloniatis}
\address{Special Research Centre for the Subatomic Structure of Matter,
University of Adelaide, 
South Australia 5005, Australia}
\author{Sergei N. Nedelko } 
\address{Institut f\"ur Theoretische Physik III
Universit\"at Erlangen - N\"urnberg,
Staudtstra{\ss}e 7
D-91058 Erlangen, Germany,
and Bogoliubov Laboratory of Theoretical Physics, JINR,
141980 Dubna, Russia}

\date{15 June, 2001}
\maketitle

\begin{abstract} 
We suggest that clusters or domains of
topological charge and action density
occur in the QCD vacuum as an effect of singularities in gauge
fields and can simultaneously lead to confinement and chiral symmetry breaking.
The string constant, condensates
and topological susceptibility are estimated within a simplified model
of hyperspherical domains
with interiors of constant field strength with reasonable values obtained. 
Propagators of dynamical quarks and gluons
have compact support in configuration space, thus having entire
Fourier transforms, which gives rise to their confinement.
\end{abstract}

]

Since Gribov and Singer \cite{GrSi}, evidence has mounted that 
singularities in gauge vector potentials are intrinsically 
unavoidable in nonabelian theories.  Several scenarios of
confinement  based on "condensation"
of singular confirations such as monopoles and
vortices, typical for  
the maximal abelian, Laplacian or axial gauges, 
are the subject of intensive analytical and numerical 
lattice studies today \cite{monopoles}. 
These focus mainly on the singular fields themselves,
their emergence from gauge fixing, their topological
properties and role in the QCD functional integral. 

Here we emphasise a less studied effect due to
singular gauge fields, namely their restrictive
influence on fluctuations in the vicinity of
singularities~\cite{lenz1,lenz2}. The subtleties of separating  
fields into regular and singular parts and the behaviour of regular
fields at the singularities are irrelevant if one could calculate 
the QCD functional integral ``exactly''. 
But these issues   
become crucial if one  undertakes approximations~\cite{lenz2}. 
In gauge invariant quantities singularities due to  
ambiguities in gauge fixing should not occur. 
This can happen in the action either due to cancellations between 
derivative and commutor
parts in the field strength if the singularity is topologically nontrivial
(monopole or vortex) or due to finitness of both terms separately
for topologically trivial singularities (domain walls,  
which, in general, can also appear~\cite{ford}).
The cancellation of singularities in the action density
can be prohibited by fluctuations around them.
Thus finiteness of the action 
implies specific constraints on fluctuations as is 
discussed in Ref.~\cite{lenz2} where 
Polyakov gauge monopoles are considered. 
 
We formulate a model partition function which 
incorporates singularities in gauge fields  
effectively via their restrictive effect on fluctuations. 
We assume that singularities are present in general
in gauge potentials, and in their vicinity  
one can divide an arbitrary field $A$ into 
singular pure gauge $S$ and regular $Q$ parts: $ A_\mu=S_\mu + Q_\mu. $
The field strength for pure gauge $S$ vanishes, but the field strength
and the action density for $A$ is singular unless the field $Q$ satisfies 
certain conditions in the vicinity of the singularities in $S$. 
To be explicit, at the cost of generality,  
we further assume that singularities in vector 
potentials are concentrated on hypersurfaces $\partial V_j$ ($j=1,\dots,N$) 
in Euclidean space of volume $V$, in the vicinity of which
gauge fields can be divided as above into a sum of a singular pure gauge  
$S^{(j)}_{\mu}$ and regular fluctuation part 
$Q^{(j)}_{\mu}$, with a colour vector $n_j^a$ associated with
$S^{(j)}$.  For such fields  
to have finite action the fluctuations charged with 
respect to $n_j$ must obey specific conditions on $\partial V_j$. 
The interiors of these regions thus constitute ``domains'' $V_j$.
Demanding finiteness of the classical action density, one arrives at 
\begin{eqnarray}
\label{bc}
\breve n_j Q^{(j)}_\mu = 0, \
\psi=-i\!\not\!\eta^j e^{i\alpha_j\gamma_5}\psi,
\
\bar\psi=\bar\psi i\!\not\!\eta^j e^{-i\alpha_j\gamma_5},
\end{eqnarray}
for $\ x\in\partial V_j$,
with the adjoint matrix $\breve n_j=T^an_j^a$ in the condition
for gluons, and a bag-like boundary condition for quarks,
$\eta^j_\mu(x)$ being a unit vector normal to $\partial V_j$.

Equations~(\ref{bc}) indicate that 
gauge modes neutral with respect to $n^a_j$ are not restricted
and provide for interactions between domains.  
In a given domain $V_j$ the effect of fluctuations in 
the rest of the system is manifested by an external gauge field 
$B_{j\mu}^a$ neutral with respect to $n_j^a$. This enables an
approximation in which domains are treated as decoupled but,
simultaneously, a compensating mean field is introduced in their interiors.
The model becomes analytically tractable if we consider
spherical domains with fixed radius $R$ and approximate the 
mean field in $V_j$ by a covariantly constant (anti-)self-dual 
configuration with the field strength
\begin{eqnarray}
\label{meanfield}
&&\hat {\cal B}_{\mu\nu}^{(j)}
=\hat n^{(j)}B^{(j)}_{\mu\nu},
\
\tilde B^{(j)}_{\mu\nu}=\pm B^{(j)}_{\mu\nu},
\
B^{(j)}_{\mu\nu}B^{(j)}_{\rho\nu}=B^2\delta_{\mu\rho},
\nonumber
\\
&&\hat n^{(j)}=t^3\cos\xi_j+t^8\sin\xi_j
\ , \  \xi_j\in\{(2k+1)\pi/6\}_{ k=0}^5,
\end{eqnarray}
where the parameter $B={\rm const}$ is the same for all domains
and the constant matrix $n_j^at^a$ belongs to the Cartan subalgebra.
Note that there is no source for this field on the boundary
and therefore it should be treated as strictly homogeneous in all further
calculations. The homogeneity itself appears as an approximation.

A self-consistent mean field approach requires 
calculation of the effective action as 
a functional of the mean field and characteristic functions 
of the domains. Its minima would give information about 
mean field character, shape and typical domain size. Needless to say that 
this problem has yet to be formulated. Nonetheless in  Eqs.~(\ref{meanfield})
we have already assumed the effective action favouring nonzero mean field 
strength parameter $B$ and finite typical size $R$. With this 
and arbitrary constant mean field, it can be shown
that the effective action for a domain 
exhibits twelve degenerate
discrete minima corresponding to (anti-)self-dual configurations 
and six values
(for $SU(3)$) of the angle $\xi$ associated with the Weyl group. 
There is also a 
degeneracy in the orientation of the chromomagnetic field.
The value $\xi_0=\pi/6$ is specific for an {\it ansatz} with the 
effective action polynomial in ${\rm Tr}\hat {\cal B}^k$, 
but the period $\pi/3$ is universal.
Since the volume of the domain is finite the degenerate minima 
do not correspond to thermodynamical phases and 
have to be summed in the partition function.   
The partition function for the model is defined as
\begin{eqnarray}
&&{\cal Z}={\cal N}\lim_{V,N}
\prod\limits_{i=1}^N\int\limits_V\frac{d^4z_i}{V}
\int\limits_{\Sigma}d\sigma_i
\int\limits_{{\cal F}^i_Q} {\cal D}Q^i
\int\limits_{{\cal F}^i_\psi}{\cal D}\psi_i 
{\cal D}\bar \psi_i\times
\nonumber\\
&&
\delta[D(\breve{\cal B}^{(i)})Q^{(i)}]
\Delta_{\rm FP}[\breve{\cal B}^{(i)},Q^{(i)}]
e^{
- S_{V_i}^{\rm QCD}
\left[Q^{(i)}+{\cal B}^{(i)},\psi^{(i)},\bar\psi^{(i)}\right]
},
\nonumber
\end{eqnarray}
where the thermodynamic limit assumes $V,N\to\infty$  
with the density $v^{-1}=N/V$ taken finite. 
The fields $Q^{(i)}$, $\psi_i$ and $\bar\psi_i$ are subject to
boundary conditions Eq.(\ref{bc}), in which
the original singularities are effectively encoded.
Interaction between the original domains is
substituted by the mean field.  A background gauge condition is imposed.
The integration measure $d\sigma_i$ is 
\begin{eqnarray}
\label{measure}
&&\int\limits_{\Sigma}d\sigma_i\dots=\frac{1}{48\pi^2}
\int\limits_0^{2\pi}d\alpha_i
\int\limits_0^{2\pi}d\varphi_i\int\limits_0^\pi d\theta_i\sin\theta_i
\times
\\
&&\int\limits_0^\pi d\omega_i\sum\limits_{k=0,1}\delta(\omega_i-\pi k)
\int\limits_0^{2\pi} d\xi_i
\sum\limits_{l=0}^{5}\delta\left(\xi_i-(2l+1)\frac{\pi}{6}\right)
\dots .
\nonumber
\end{eqnarray}
Here $\varphi_i$ and $\theta_i$ are spherical angles of
the chromomagnetic field,
$\omega_i$ is the angle between the chromomagnetic and
chromoelectric fields, $\xi_i$ is the angle in the colour
matrix $\hat n_i$, 
$\alpha_i$ is the chiral angle 
and $z_i$ is the centre of the domain $V_i$.

This partition function describes a statistical system 
of density $v^{-1}$ composed of noninteracting clusters, each of which is
characterised by a set of internal parameters and
internal dynamics represented by the fluctuation fields.
Correlation functions can be calculated 
taking the mean field into account
explicitly and decomposing over the fluctuations.
First of all we consider vacuum characteristics of the system
in the  zeroth order of this expansion.

The connected $n-$point correlator of the mean field
strength tensor is found to be  
\begin{eqnarray}
&&\langle B^{a_1}_{\mu_1\nu_1}(x_1)\dots B^{a_n}_{\mu_n\nu_n}(x_n) \rangle
= B^{n} t^{a_1\dots a_n}_{\mu_1,\dots,\nu_n}
\Xi_n(x_1,\dots,x_n),
\nonumber\\
&& B^{a}_{\mu\nu}(x)
=\sum_j^N n^{(j)a}B^{(j)}_{\mu\nu}\theta(1-(x-z_j)^2/R^2),
\nonumber\\
&&t^{a_1\dots a_n}_{\mu_1,\dots,\nu_n}=B^{-n}
\int d\sigma_j
n^{(j)a_1}\dots n^{(j)a_n}B^{(j)}_{\mu_1\nu_1}\dots B^{(j)}_{\mu_n\nu_n},
\nonumber\\
&&\Xi_n=\frac{1}{v}\int d^4z
\theta\left(1-\frac{(x_1-z)^2}{R^2}\right)\dots
\theta\left(1-\frac{(x_n-z)^2}{R^2}\right),
\nonumber
\end{eqnarray}
by explicit calculation using the measure, Eq.~(\ref{measure}). 
Translation-invariant functions $\Xi_n(x_1,\dots,x_n)$
are equal to the volume of the overlap region for $n$ 
hyperspheres of radius $R$ and centres ($x_1,\dots,x_n$),
normalised to the volume of a single hypersphere
$v=\pi^2R^4/2$. They are continuous and vanish if $|x_i-x_j|\ge 2R$: 
correlations in the background field have finite range
$2R$ and have no particle interpretation. The statistical ensemble 
of background fields is non-Gaussian since all connected correlators
are independent and cannot be reduced to the two-point correlators.
The simplest application of this is 
a gluon condensate 
$$g^2 \langle F^a_{\mu\nu}(x)F^a_{\mu\nu}(x)\rangle=4B^2.$$

A significant vacuum parameter for the resolution of the $U_A(1)$ problem
is the topological susceptibility \cite{Cre77}. First we consider    
the topological charge density
for colour $SU(3)$ in this approximation 
\begin{eqnarray}
Q(x)={{g^2}\over {32 \pi^2}} \tilde F F
={{B^2}\over {8 \pi^2}} \sum_{j=1}^N\theta[1-(x-z_j)^2/R^2]\cos\omega_j,
\nonumber
\end{eqnarray} 
where $\omega_j\in\{0,\pi\}$ depends on the duality of the $j$-th domain.
The topological charge is additive 
$$
Q=\int_V d^4x Q(x)= q(N_+ - N_-), \  -Nq\le Q\le Nq,
$$ 
where $q=B^2R^4/16$ is a `unit' topological charge,
namely the absolute value of the topological charge of a single
domain,  and $N_{+}$ $(N_-)$ is the number of 
domains with (anti-)self-dual field, $N=N_+ + N_-$. 
The probability distribution for topological charge reads
$$
{\cal P}_N(Q)=\frac{{\cal N}_N(Q)}{{\cal N}_N}=
\frac{N!}{2^N\left(N/2-Q/2q\right)!\left(N/2+Q/2q\right)!},
$$
where ${\cal N}_N(Q)$ is the number of configurations with a given charge
and ${\cal N}_N$ is the total number of configurations.
The distribution is symmetric about $Q=0$, where it has a maximum
for $N$ even.  For $N$ odd the maximum is at $Q=\pm q$. 
Averaged topological charge is zero. 
The topological susceptibility
is then determined by the two-point correlator of topological charge density:
\begin{eqnarray}
\chi = \int d^4x \langle Q(x) Q(0) \rangle
 = {{B^4 R^4} \over {128 \pi^2}}.
\nonumber
\end{eqnarray}
 
Static quark confinement is examined via calculation of the Wilson loop
\begin{eqnarray}
W(L)=\lim_{N\to\infty}\prod\limits_{j=1}^N\int_V\frac{d^4z_j}{V}
\int\frac{ d\sigma_j}{N_c}{\rm Tr}
e^{i\int_{S_L}d\sigma_{\mu\nu}(x)\hat B_{\mu\nu}(x)}.
\nonumber
\end{eqnarray}
Path ordering here is unnecessary as the matrices 
$\hat n^k$ belong to the Cartan subalgebra.
For computational convenience  
we consider a circular contour in the $(x_3,x_4)$
plane of radius $L$ with centre at the origin.  
Calculation of the colour trace, integrating over spatial 
orientations of the mean field and positions of the domains
and then taking the thermodynamic limit ($N\to\infty$, 
$v=V/N=\pi^2R^4/2$),
gives an area law for a large Wilson loop $L\gg R$  
\begin{eqnarray}
\label{sig-su3}
&&W(L)=e^{\sigma \pi L^2 + O(L)}, 
\
\sigma=Bf(\pi BR^2), 
\nonumber\\ 
&&f(z)=\frac{2}{3z}
\left(3-
\frac{\sqrt{3}}{2z}\int\limits_0^{\frac{2z}{\sqrt{3}}}\frac{dx}{x}\sin x
-
\frac{2\sqrt{3}}{z}\int\limits_0^{\frac{z}{\sqrt{3}}}\frac{dx}{x}\sin x
\right).
\nonumber
\end{eqnarray}
The function $f$ is positive for $z>0$ and has a maximum for $z= 1.55\pi$.
We choose this maximum to estimate the model parameters 
by fitting the string constant to the lattice result,
\begin{eqnarray}
\label{par-val}
\sqrt{B}=947{\rm MeV}, \ R^{-1}=760 {\rm MeV},
\end{eqnarray}
with unit charge $q=0.15$, density  $v^{-1}=42.3 {\rm fm}^{-4}$ 
and the ``observable'' 
gluonic parameters of the vacuum
\begin{eqnarray}
\label{res1}
&&\sqrt{\sigma}=420 \ {\rm MeV}, \ \chi=(197 \ {\rm MeV})^4,
\nonumber\\ 
&&(\alpha_s/\pi)\langle F^2\rangle=0.081 \ {\rm GeV}^4.
\end{eqnarray}
This result indicates high density and strong background fields
in the system. 
There is no separation of scales, $\sqrt{B}R\approx 1$. 
Neither large domains nor stochasticity   
of background fields are seen here
which {\it a posteriori} justifies the mean
field averaging prescription in the partition 
function.  The high density ensures area law dominance  
already at distances $2L\approx 1.5-2\ {\rm fm}$. 
Taylor-expanding the integrand in the Wilson loop integral
would reveal all $n$-point correlation functions of the
background field, and arguments about 
the importance of a fast decay of correlators
for static confinement \cite{DoSim} would then be seen to apply here.

Due to averaging over self- and anti-self-dual 
configurations and angles $\alpha_i$
there is no explicit violation of parity and chiral symmetry 
in the partition function.
A spontaneous breaking can be tested by estimating the quark condensate.
To first order in fluctuations this  
requires calculation of the quark propagator with the condition
\begin{eqnarray}
&&i\!\not\!\eta(x)e^{i\alpha\gamma_5}S(x,y)=-S(x,y), \ (x-z)^2=R^2.
\nonumber
\end{eqnarray} 
An analogous condition holds for $(y-z)^2=R^2$. Substitution 
(with the upper (lower) sign for (anti-)self-dual domains)
\begin{eqnarray}
&&S=(i\!\not\!D+m)
[P_\pm{\cal H}_0+P_\mp O_+{\cal H}_1 + P_\mp O_-{\cal H}_{-1}]
\label{q-pr1},
\nonumber\\
&&O_\pm=
[1\pm \hat n\vec\Sigma\vec B/|\hat n| B]/2, 
\ P_\pm=[1\pm\gamma_5]/2, 
\nonumber
\end{eqnarray}
leads to equations for the scalar functions 
 ${\cal H}_{\zeta}$:
\begin{eqnarray}
(-D^2+m^2+2 \zeta \hat B){\cal H}_{\zeta}(x,y)=\delta(x,y),
\nonumber
\end{eqnarray}
where $\hat B=|\hat n| B$.
Solutions vanishing at infinity would give a Green's 
function ${\cal H}_{-1}$ divergent in the massless limit 
due to zero modes of the Dirac operator. The bag-like boundary conditions 
remove zero eigenvalues from the spectrum, and the massless limit is regular.

To render our calculation transparent we choose  
$y=z=0$ and calculate the quark condensate at the domain centre.  
Then the general solutions 
for scalar Green's functions take the form ($\mu_{\zeta}=m^2/2\hat B + \zeta$)
\begin{eqnarray}
&&{\cal H}_{\zeta}=\Delta(x^2|\mu_{\zeta})+
C_{\zeta}e^{-\hat Bx^2/4}M(1+\mu_{\zeta},2,\hat Bx^2/2),
\nonumber
\end{eqnarray}
where $\Delta(x^2|\mu)$ is the vanishing at infinity scalar propagator 
for mass $2\hat B\mu$, and the second term 
is a homogeneous solution regular at $x^2=0$  
expressed via the confluent hypergeometric function.
The constants $C_\zeta$ can be fit to implement the boundary condition.
The terms  $m{\cal H}_0$ and $m{\cal H}_1$ vanish in the massless limit
and do not contribute to the condensate.
The nontrivial contribution comes from the term $m{\cal H}_{-1}$.
The bag-like conditions imply that on the boundary
${\cal H}_{-1}$ satisfies a mixed condition with $f'=df/d|x|$
and the sign ($-$)$+$ corresponding to (anti-)self-dual domain,
\begin{eqnarray}
&&2e^{\mp i\alpha}m {\cal H}_{-1} = -2{\cal H}_{-1}' 
- \hat BR^2{\cal H}_{-1},
\nonumber
\end{eqnarray} 
which leads to the relations
\begin{eqnarray}
&&\lim_{m\to 0}m{\cal H}_{-1}=
\frac{e^{\pm i\alpha}}{2\pi^2R^3}F(\hat BR^2/2)e^{-\hat Bx^2/4},
\nonumber\\
&&{\rm Tr}S(0,0)=\frac{e^{\pm i\alpha}}{2\pi^2R^3}\sum\limits_{|\hat n|}
F(\hat BR^2/2),
\nonumber\\
&&F(z)=e^z-z-1+\frac{z^2}{4}\int\limits_0^\infty
\frac{dte^{2t-z({\rm coth}t-1)/2}}{{\rm sinh}^2t} 
({\rm coth}t-1).
\nonumber
\end{eqnarray}
Note that the term in the propagator with nonzero trace 
has definite chirality correlated with the duality of domain and
is proportional to the zero mode
of the Dirac operator: $\!\not\!DP_\mp O_{-}\exp(-\hat Bx^2/4)=0$.

Averaging this result over $\alpha$ and 
(anti-)self-dual configurations and taking into 
account the $\alpha$-dependence of the quark determinant~\cite{wipf} 
$$
{\rm det}S^{-1}
\propto 
\exp\{i\alpha
\int_v dx Q(x)\}
\approx \exp\{\pm i q\alpha\},
$$
with $q$ being the unit topological charge, we get a finite result
\begin{eqnarray}
\langle \bar\psi \psi\rangle= 
-\frac{q}{2\pi^2R^3(1+q)}\sum\limits_{|\hat n|}
F(\hat BR^2/2)= -(228 {\rm MeV})^3,
\nonumber
\end{eqnarray}
for $B$ and $R$ as determined above, Eq.~(\ref{par-val}).
Obviously the condensate vanishes for zero topological charge $q$.     
Final conclusions require complete calculation
of the condensate 
and detailed analysis of the eigenvalue problem.

Confinement of dynamical quarks and gluons 
can be studied in this context   
via the analytical properties of their propagators. 
As has been partly demonstrated for quarks, 
the propagators of quark and gluon fluctuations
can be analytically calculated by reduction 
to the scalar problem, essentially that  
of a four-dimensional harmonic oscillator with  
total angular momentum coupled to the external field. 
The general solution is given by decomposition over hyperspherical harmonics.
Qualitatively, the boundary conditions mean  
$x-$space propagators of charged fields are defined in
regions of finite support where they have standard ultraviolet
singularities. So their Fourier transforms 
are entire functions in the complex momentum plane\cite{NKYM01}.
This manifests confinement of dynamical fields\cite{leutw,efi}. 
A known consequence of entire propagators is a Regge spectrum of 
relativistic bound states\cite{efi}.

The model displays the most essential features of QCD and,
being 
analytically tractable, provides a framework for phenomenological applications.
The first test in this direction should be
a calculation of the spectrum of bound states via the
Bethe-Salpeter framework, or within a hadronisation
scheme analogous to \cite{efi}.
The model preserves
the structure of a relativistic quantum field theory,
albeit a nonlocal one~\cite{efi1}.
Certainly one has to be convinced that the qualitative arguments
underlying the model are formally legitimate.  
Associating a colour direction with the
boundary is a restrictive assumption. Somewhat akin to
"abelian projection", it singles out a class
of abelian mean fields, and ``neutral'' gluons
are not confined.
Incorporating topologically nontrivial singularities can
improve this aspect of the model.

Clustering of action and topological charge density with
dominance by (anti-)self-dual fields in  
the QCD vacuum is evident in lattice cooling algorithms,
which, by design, iterate towards (anti-)instantons~\cite{Tep99}.
Alternately, Horv{\'a}th et al. \cite{Hor01} suggested studying duality 
of clusters via the chirality of fermionic eigenmodes. 
Used with overlap fermions 
this shows strong correlation between clusters of topological
charge and locations where several low eigenvalue fermionic modes are
largest; the modes are also chiral  
\cite{DeGHas01}. This is regarded as evidence for instantons. 
Such correlation between chirality of fermionc modes and duality
of domains is also evident in the model discussed above.
Our estimates suggest that
formation of clusters, predominantly (anti-)self-dual and 
with average size $2R\approx 0.5 {\rm fm}$,  
can have purely quantum origin
whose explanation could require reference 
to the existence of obstructions in gauge fixing
rather than to the quasi-classical limit.

\smallskip

\noindent
{\bf Acknowledgements}\\
ACK is supported by the Australian Research Council.
We are grateful to Frieder Lenz 
for fruitful intensive discussions and encouraging criticism,
and  to Garii Efimov, Jan Pawlowski and 
to Lorenz von Smekal for valuable comments.

\end{document}